\documentclass[aps,floats,prl,twocolumn]{revtex4}
\usepackage{amsfonts}
\usepackage{amsmath}
\usepackage{amssymb}
\usepackage[final,dvips]{graphicx}
\begin{document}
\title{Why magnesium diboride is not described by anisotropic Ginzburg-Landau theory}
\author{A. E. Koshelev}
\affiliation{Materials Science Division, Argonne National Laboratory,Argonne, Illinois 60439}
\author{A. A. Golubov}
\affiliation{Faculty of Science and Technology, University of
Twente, 7500 AE Enschede, The Netherlands} \keywords{} \pacs{}
\date{\today}

\begin{abstract}
It is well established that the superconductivity in the recently
discovered superconducting compound MgB$_{2}$ resides in the
quasi-two-dimensional band ($\sigma$-band) and three-dimensional
band ($\pi$-band). We demonstrate that, due to such band
structure, the anisotropic Ginzburg-Landau theory practically does
not have region of applicability, because gradient expansion in
the $c$ direction breaks down. In the case of dirty $\pi$-band we
derive the simplest equations which describe properties of such
superconductors near $T_{c}$ and explore some consequences of
these equations.

\end{abstract}
\maketitle

Ginzburg-Landau (GL) theory is the most powerful and widely used
phenomenological theory of superconductivity (see, e.\ g., Refs.\
\onlinecite{GL,Tinkham}). It describes practically all known
superconductors in the vicinity of transition temperature. GL
theory is fully microscopically justified and all its parameters
can be derived from the microscopic BCS theory \cite{Gorkov}. GL
theory provides the basis for such elaborated fields as vortex
physics \cite{Abrikosov} and the theory of fluctuation phenomena
\cite{Tinkham}.

The recently discovered superconductor MgB$_{2}$ \cite{Akimitsu}
gives an example of a superconductor which \emph{is not described
by the anisotropic GL theory}. This very unusual feature is a
consequence of a specific band structure of this compound. It is
reliably established that superconductivity in MgB$_{2}$ resides
in two families of bands: strongly superconducting
quasi-two-dimensional $\sigma$-bands and weakly superconducting
three-dimensional $\pi$-bands (see, e.g.,
Ref.\ \cite{PhysCSpecial}). Both bands are characterized by their
intrinsic coherence lengths, and the $c$ axis coherence length in
the $\sigma$-band is much smaller than $c$ axis coherence length
in the $\pi$-band. Typically, the strong band forces the order
parameter in the weak band to change in the $c$ direction at
distances smaller than the intrinsic $c$ axis coherence length in
this band. This means that almost in the whole temperature range
the effective coherence length in the $c$ direction, $\xi_{z}(T)$,
is smaller than the intrinsic coherence length in the $\pi$-band,
$\xi_{\pi,z}$. The crossover to the GL region takes place only
when $\xi_{z}(T)$ exceeds $\xi_{\pi,z}$, which occurs in the very
close vicinity of $T_{c}$. In this narrow region the $\pi$-band
strongly increases the $c$ axis coherence length. Beyond the
narrow region, the variations of the order parameter in the $c$
directions are not described by the anisotropic GL theory.
Important consequencies of GL theory breakdown are the strong
temperature dependence of the $H_{c2}$ anisotropy
\cite{Hc2Clean,Hc2Exp,GurevichPRB03,GK-Hc2} and strong deviations
of the $H_{c2}$ angular dependence from the simple "effective
mass" law \cite{GK-Hc2,Rydh03}.

Obviously, the breakdown of the anisotropic GL theory has numerous
consequences and it would be desirable (i) to trace the reason of
this breakdown and (ii) to derive the simplest model, which
replaces the GL model near $T_c$. This paper addresses these
issues. For illustration, we use the simplest microscopic model,
multiband generalization of the Usadel theory, describing a dirty
two-band superconductor with weak interband scattering
\cite{KG,GurevichPRB03}. However, the main conclusions are very
general and do not depend much on the intraband scattering
strength. In the model we use the GL expansion for the
$\sigma$-band and keep the microscopic description for the
$\pi$-band, i.e., only "dirtyness" of the $\pi$-band is essential
for a particular form of equations.

We consider a dirty two-band superconductor with weak interband
scattering. Such superconductor is described by Usadel equations
for the impurity averaged normal and anomalous Green's functions,
$G_{\alpha}$ and $F_{\alpha}$, $G_{\alpha}^{2}\!+\!\left\vert
F_{\alpha}\right\vert ^{2}\!=\!1$, and the pair potentials
$\Delta_{\alpha}$,
\begin{equation}
\omega F_{\alpha}\!-\!\sum\nolimits_{j}\frac{\mathcal{D}_{\alpha,j}}{2}\left[
G_{\alpha}D_{j}^{2}F_{\alpha}\!-\!F_{\alpha}\nabla_{j}^{2}G_{\alpha}\right]
=\Delta_{\alpha}G_{\alpha}\label{Falpha}
\end{equation}
where $\alpha=1,2$ is the band index, $j=x,y,z$ is the coordinate
index, $D_{j}\equiv\nabla_{j}\!-\!(2\pi i/\Phi_{0})A_{j}$,
$\mathcal{D} _{\alpha,j}$ are diffusion constants, and
$\omega=2\pi T(s+1/2)$ are Matsubara frequencies. Bearing in mind
the application to MgB$_{2}$, in our notations index 1 corresponds
to $\sigma$-bands and index 2 to $\pi$-bands,
$\mathcal{D}_{1,j}\equiv \mathcal{D}_{\sigma,j}$ and
$\mathcal{D}_{2,j}\equiv\mathcal{D}_{\pi,j}$. All bands are
isotropic in the $xy$ plane, $\mathcal{D}_{\alpha,x}=\mathcal{D}
_{\alpha,y}$ and anisotropic in the $xz$ plane with the anisotropy
ratios
$\gamma_{\alpha}=\sqrt{\mathcal{D}_{\alpha,x}/\mathcal{D}_{\alpha,z}}$.
Self-consistency conditions can be written as \cite{KG}
\begin{subequations}
\label{SelfCons}
\begin{align}
W_{1}\Delta_{1}\!-\!W_{12}\Delta_{2} &  \!=\!2\pi T\sum_{\omega>0}\left(
F_{1}\!-\!\frac{\Delta_{1}}{\omega}\right)  \!+\!\Delta_{1}\ln\frac{1}
{t}\label{SelfCons1}\\
-W_{21}\Delta_{1}\!+\!W_{2}\Delta_{2} &  \!=\!2\pi T\sum_{\omega>0}\left(
F_{2}\!-\!\frac{\Delta_{2}}{\omega}\right)  \!+\!\Delta_{2}\ln\frac{1}
{t}\label{SelfCons2}
\end{align}
\end{subequations}
where $t\equiv T/T_{c}$ and the matrix $W_{\alpha\beta}$ is
related to the matrix of coupling constants
$\Lambda_{\alpha\beta}$ as
\begin{align*}
W_{1} &  \!=\!\frac{-\Lambda_{-}\!+\!\sqrt{\Lambda_{-}^{2}\!+\!\Lambda
_{12}\Lambda_{21}}}{\mathrm{Det}},\ W_{2}\!=\!\frac{\Lambda_{-}\!+\!\sqrt
{\Lambda_{-}^{2}\!+\!\Lambda_{12}\Lambda_{21}}}{\mathrm{Det}},\\
W_{12} &  =\Lambda_{12}/\mathrm{Det},\ W_{21}=\Lambda_{21}/\mathrm{Det,}
\end{align*}
$\Lambda_{-}\equiv (\Lambda_{11}-\Lambda_{22})/2,$
$\mathrm{Det}\equiv \Lambda_{11}
\Lambda_{22}-\Lambda_{12}\Lambda_{21}$, $W_{1}W_{2}=W_{12}W_{21}$.
The supercurrent components are given by
\begin{equation}
j_{j}=4\pi eT\sum_{\alpha}\sum_{\omega>0}N_{\alpha}\mathcal{D}_{\alpha
,j}\operatorname{Im}\left[  F_{\alpha}^{\ast}D_{j}F_{\alpha}\right]  ,
\end{equation}
where $N_{\alpha}$ are the partial densities of states

We start with the derivation of the GL equations from the Usadel
equations in the close vicinity of $T_{c}$ following a standard
route. In the lowest approximation $G_{\alpha}^{(0)}=1$ and
$F_{\alpha}^{(0)}\approx\Delta_{\alpha }/\omega$. When
$\Delta_{\alpha}$ are small and change slowly in space (the exact
criterion will be derived below) one can keep only the leading
nonlinear and gradient corrections
\begin{equation}
F_{\alpha}\approx\frac{\Delta_{\alpha}}{\omega}-\frac{\Delta_{\alpha}^{3}
}{2\omega^{3}}+\sum\nolimits_{j}\frac{\mathcal{D}_{\alpha, j}}{2\omega^{2}
}D_{j}^{2} \Delta_{\alpha}.\label{Expansion}
\end{equation}
Substituting this expansion into the self-consistency conditions
(\ref{SelfCons}),
we obtain coupled GL equations for two gap
parameters\cite{cleanGL}
\begin{subequations}
\label{couplGL}
\begin{align}
W_{1}\Delta_{1}-W_{12}\Delta_{2}  &  =\sum\nolimits_{j}\xi_{1, j}^{2}D_{j}
^{2}\Delta_{1} -b\Delta_{1}^{3}+\tau\Delta_{1}\label{GL1}\\
-W_{21}\Delta_{1}+W_{2}\Delta_{2}  &  =\sum\nolimits_{j}\xi_{2, j}^{2}
D_{j}^{2}\Delta_{2} -b\Delta_{2}^{3}+\tau\Delta_{2}\label{GL2}
\end{align}
with $\xi_{\alpha, j}^{2}=(\pi/8T)\mathcal{D}_{\alpha, j}$,
$b=7\zeta(3)/(8\pi^{2}T^{2})$, and
$\tau=\ln(1/t)\approx(T_{c}-T)/T_{c}$.

Near $T_{c}$ the right hand sides of Eqs.\ (\ref{couplGL}a,b) are small. This
allows us to reduce Eqs.\ (\ref{couplGL}a,b) to a single GL equation by
looking for solution for $\Delta_{2}$ in the form
\end{subequations}
\begin{equation}
\Delta_{2}\approx\frac{W_{21}}{W_{2}}\Delta_{1}+\delta_{2}\text{,}
\label{D2Present}
\end{equation}
From Eqs.\ (\ref{couplGL}a,b) we obtain
\begin{subequations}
\begin{align}
-W_{12}\delta_{2} & =\sum\nolimits_{j}\xi_{1, j}^{2}D_{j}^{2}\Delta
_{1}-b\Delta_{1}^{3} +\tau\Delta_{1}\\
W_{2}\delta_{2} & =\sum\nolimits_{j}\xi_{2,
j}^{2}D_{j}^{2}\Delta_{2} -b\Delta_{2}^{3}+\tau\Delta_{2}.
\end{align}
The second equation indicates that $\delta_{2}$ is a small correction,
$\delta_{2}\ll\Delta_{2}$, and one can use $\Delta_{2}\approx\left(
W_{21}/W_{2}\right)  \Delta_{1}$ in the right hand side of this equation.
Excluding $\delta_{2}$ and introducing the band-averaged order parameter
\end{subequations}
\[
\Delta^{2}=\frac{W_{2}\Delta_{1}^{2}+W_{1}\Delta_{2}^{2}}{W_{2}+W_{1}}
\approx\frac{W_{12}W_{2}^{2}+W_{21}W_{1}^{2}}{W_{12}W_{2}\left(  W_{2}
+W_{1}\right)  }\Delta_{1}^{2}
\]
we finally obtain the anisotropic GL equation for $\Delta$
\begin{equation}
-\sum\nolimits_{j}\xi_{j}^{2}D_{j}^{2}\Delta+b\Delta^{3}-\tau\Delta=0,\label{anisGL}
\end{equation}
with the average coherence lengths
\[
\xi_{j}^{2}=\frac{W_{2}\xi_{1, j}^{2}+W_{1}\xi_{2, j}^{2}}{W_{2}+W_{1}}.
\]

For the supercurrent, using relation $W_{21}/W_{12}=\Lambda_{21}/\Lambda
_{12}=N_{1}/N_{2}$, we derive
\begin{equation}
j_{j}\approx4eNP\xi_{j}^{2}\operatorname{Im}\left[  \Delta^{\ast}D_{j}
\Delta\right] \label{GLsupercur}
\end{equation}
with $N\equiv N_{1}+N_{2}$ and
\[
P\equiv\frac{N_{1}N_{2}\left(  W_{2}+W_{1}\right)  ^{2}}{\left(
N_{1} +N_{2}\right)  \left(  N_{2}W_{2}^{2}+N_{1}W_{1}^{2}\right)
}.
\]
From Eqs.\ (\ref{anisGL}) and (\ref{GLsupercur}) we derive the
components of the London penetration depth
\begin{equation}
\lambda_{j}^{-2}\approx\frac{32\pi^{2}eNP\xi_{j}^{2}\tau}{c\Phi_{0}
b}.\nonumber
\end{equation}
For the parameters of MgB$_{2}$, $W_{1}\ll W_{2}$,
$\xi_{1,z}\ll\xi_{2,z}$, $\xi_{1,x}\sim\xi_{2,x}$, the dominating
effect of the $\pi$-band is the renormalization of the $c$ axis
lengths
\begin{align}
\xi_{z}^{2} &  \approx\xi_{1,z}^{2}+S_{12}\xi_{2,z}^{2},\label{xiGLapprox}\\
\lambda_{z}^{-2} &  \approx\frac{32\pi^{2}e\tau N_{1}}{c\Phi_{0}b}\left(
\xi_{1,z}^{2}+S_{12}\xi_{2,z}^{2}\right)  .\label{LambdaGLApprox}
\end{align}
with $S_{12}\equiv W_{1}/W_{2}\ll1$. The influence of the $\pi$-band on
properties not related with the variations of the order parameter along the
$c$ axis are weak and can be treated perturbatively.

We obtain now the applicability criterion for the GL expansion. The gradient
expansion is justified if $-\xi_{\alpha, j}^{2}\nabla_{j}^{2}\Delta_{\alpha
}<\Delta_{\alpha}$ for all $\alpha$ and $i$. Because a typical scale of the
spatial variations is the temperature-dependent GL coherence length $\xi
_{j}(T)$, this condition simply means
\begin{equation}
\xi_{j}(T)>\xi_{\alpha, j}
\end{equation}
The most restraining inequality is the one for $\alpha=2$ and
$i=z$ which gives
\begin{equation}
(T_{c}-T)/T_{c}<\xi_{1,z}^{2}/\xi_{2,z}^{2}+S_{12}\label{GLcriterion}
\end{equation}
Because $\xi_{1,z}\ll\xi_{2,z}$ and $S_{12}\ll1$, the
applicability of the GL approach is limited to extremely narrow
temperature range near $T_{c}$, i.e., the situation is very
different from conventional superconductors. For parameters of
MgB$_{2}$ this condition implies $\left(  T_{c}-T\right)
/T_{c}\ll0.05$. On the other hand, near $T_{c}$ the fluctuation
effects become important. This means that the mean-field GL theory
practically does not have a region of applicability.

We derive now the simplest theory which replaces the GL theory in
the conventional GL region $\left(  T_{c}-T\right)  /T_{c}<1$. As
the gradient expansion actually breaks down only for the
$\pi$-band, in the vicinity of $T_{c}$ we can proceed with the
expansion (\ref{Expansion}) for the $\sigma $-band, $\alpha=1$.
Substituting this expansion into the self-consistency conditions,
we obtain Eq.\ (\ref{GL1}). The $\pi$-band only weakly
renormalizes the nonlinear term and we can use the linear
approximation in this band
\begin{equation}
\omega F_{2}-\sum\nolimits_{j}\frac{\mathcal{D}_{2,
j}}{2}D_{j}^{2} F_{2}=\Delta_{2}.
\end{equation}
The $\pi$-band order parameter can again be represented by Eq.\
(\ref{D2Present}) with $\delta_2$ being a small correction.
Finding this correction from Eq.\ (\ref{SelfCons2}) and
substituting it into Eq.\ (\ref{SelfCons1}) together with the GL
expansion for $F_1$, we derive coupled equations for $\Delta_1$
and reduced $\pi$-band $F$-function $f_s$, %
$f_{s}\equiv (2\pi T W_{2}/W_{21})F_{2}(\omega_s)$\cite{criterion}
\begin{subequations}
\label{minEq}
\begin{align}
-\left(  1+S_{12}\right)  \tau\Delta_{1}+b\Delta_{1}^{3}-\sum\nolimits_{j}
\xi_{1,j}^{2}D_{j}^{2}\Delta_{1} & \nonumber\\
-S_{12}\sum_{s=0}^{\infty}\left(  f_{s}-\frac{\Delta_{1}}{s+1/2}\right)   &
=0,\label{EqD1}\\
(s+1/2)f_{s}-\frac{2}{\pi^{2}}\sum\nolimits_{j}\xi_{2,j}^{2}D_{j}^{2}f_{s} &
=\Delta_{1},\label{Eq_fs}
\end{align}
\end{subequations}
and the expression for the supercurrent
\[
j_{j}\!\approx\!4eN_{1}\xi_{1,j}^{2}\!\operatorname{Im}\left[
\Delta_{1}^{\ast }D_{j}\Delta_{1}\right]
+\frac{8e}{\pi^{2}}N_{1}S_{12}\xi_{2,j}^{2}
\!\sum_{s=0}^{\infty}\operatorname{Im}\left[
f_{s}^{\ast}D_{j}f_{s}\right].
\]
These equations replace the GL
equations in the case of dirty $\pi$-band. Note that the same
equations are also valid in the case of clean $\sigma$-band but
with different definition of the coherence length $\xi_{1,j}$,
$\xi_{1,j} ^{2}=7\zeta(3)\left\langle v_{1,j}^{2}\right\rangle
/(4\pi T)^{2}$.

In the case of weak superconductivity in the $\pi$-band,
$S_{12}\ll1$, and for $\xi_{1,z}\ll$ $\xi_{2,z}$, one can neglect
the in-plane gradients in the Eq.\ (\ref{Eq_fs}) and obtain an
even simpler set of equations which describe only \emph{the
dominating strong effects}, related to inhomogeneities of the gap
parameter along the $c$ axis, and neglect small renormalizations
of the coefficients by the weak $\pi$-band
\begin{subequations}
\label{Eqs_z}
\begin{align}
-\tau\Delta_{1} &  +b\Delta_{1}^{3}-\sum\nolimits_{j}\xi_{1,j}^{2}D_{j}
^{2}\Delta_{1}\nonumber\\
&  -\!S_{12}\sum_{s=0}^{\infty}\left(  f_{s}\!-\!\frac{\Delta_{1}}
{s+1/2}\right)  \!=\!0,\label{EqD1,z}\\
&  \left(  s+1/2\right)  f_{s}-\frac{2}{\pi^{2}}\xi_{2,z}^{2}D_{z}^{2}
f_{s}=\Delta_{1},\label{Eqfs_z}\\
j_{j} &  \approx4eN_{1}\xi_{1,j}^{2}\operatorname{Im}\left[  \Delta_{1}^{\ast
}D_{j}\Delta_{1}\right] \nonumber\\
&
+\delta_{j,z}\frac{8e}{\pi^{2}}N_{1}S_{12}\xi_{2,z}^{2}\sum_{s=0}^{\infty
}\operatorname{Im}\left[  f_{s}^{\ast}D_{z}f_{s}\right]
,\label{Eq_curr_z}
\end{align}
\begin{figure}[ptb]
\begin{center}
\includegraphics[width=2.8in ]{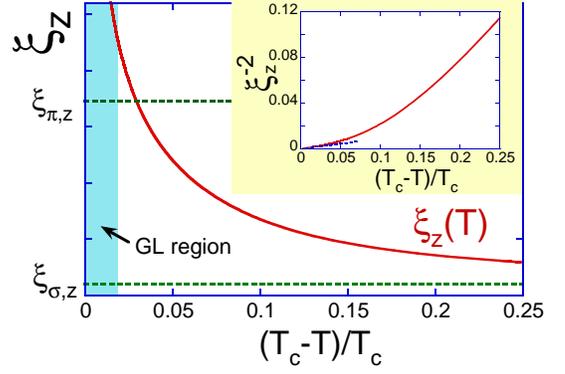}
\end{center}
\caption{Temperature dependence of the $c$ axis coherence length,
$\xi_{z}(T)$ computed from Eq.\ (\ref{xz_T}) with parameters
$S_{12}=0.034$ and $\xi _{2,z}^{2}=300\xi_{1,z}^{2}$. Marked GL
region corresponds to condition $\xi_{z}(T)>
\xi_{\pi,z}\equiv\xi_{2,z}$. The inset shows dependence $\xi
_{z}^{-2}(T)$ with the dashed line showing the linear GL
asymptotics at $T\rightarrow T_{c}$.}
\end{figure}

We explore now some consequences of these equations. To define an effective
coherence length, we consider the response of the order parameter to the weak
$z$-dependent variation of $T_{c}$, $\tau\rightarrow\tau(z)=\tau+\delta
\tau(z)$. In linear approximation with respect to $\delta\tau(z)$
Eqs.\ (\ref{EqD1,z}) and (\ref{Eqfs_z}) can be solved by Fourier transform
yielding $\Delta_{1}=\Delta_{1}^{(0)}+\delta\Delta_{1}(z)$
\end{subequations}
\begin{align*}
\delta\Delta_{1}(z) &  =\int G(z-z^{\prime})\delta\tau(z^{\prime})dz^{\prime
},\\
G(z) &
=\int\frac{dk}{2\pi}\frac{\exp(ikz)}{2\tau+\xi_{1,z}^{2}k^{2}
+S_{12}g[(2/\pi^{2})\xi_{2,z}^{2}k^{2}]},
\end{align*}
where $g(u)\equiv\psi(1/2+u)-\psi(1/2)$ and $\psi(u)$ is the
digamma-function. In contrast to the GL model, the decay of the
perturbation $\delta\Delta_{1}(z)$ is not exponential. Using the
last equation, one can introduce the effective coherence length
$\xi_{z}$, which determines the scale of spatial variations of the
order parameter in the $z$ direction,
\begin{equation}
\xi_{1,z}^{2}/\xi_{z}^{2}+S_{12}g\left[  (2/\pi^{2})\xi_{2,z}^{2}/\xi_{z}
^{2}\right]  =\tau\label{xz_T}
\end{equation}
The dependence $\xi_{z}(T)$ computed from this equation using
parameters $S_{12}\!=\!0.034$ \cite{KG} and $\xi_{2,z}^{2}\!
=\!300\xi_{1,z}^{2}$ is shown in Fig.\ 1.

\begin{figure}[ptb]
\begin{center}
\includegraphics[clip,width=3.4in ]{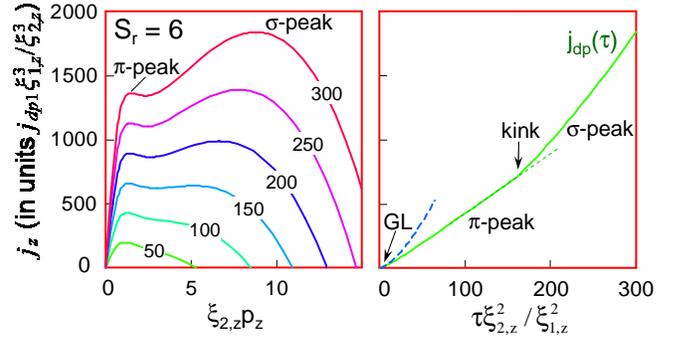}
\end{center}
\caption{\emph{Left panel:} Dependencies $j_{z}(p_z)$ at different
temperatures for $S_r=6$. The curves are marked by the reduced
temperatures $\tau \xi_{2,z}^{2}/\xi_{1,z}^{2}=50,\ldots,300$. In
the unit of the vertical axis $j_{dp1}=4eN_1 \xi_{1,z}/b$ is the
depairing current scale for the $\sigma$-band. \emph{Right panel:}
The temperature dependence of the depairing current for the same
value of $S_r$. The dashed curve shows GL dependence.}
\end{figure}
Consider the relation between the supercurrent $j_{z}$ and supermomentum
$p_{z}=\nabla_{z}\phi-(2\pi/\Phi_{0})A_{z}$, which determines the $c$ axis
London length and depairing current. From Eqs.\ (\ref{EqD1,z}) and
(\ref{Eqfs_z}) we obtain
\begin{align}
j_{z}(p_{z}) &  =4eN_{1}\Delta_{1}^{2}(p_{z})p_{z}\label{jz_pz}\\
&  \times\left(  \xi_{1,z}^{2}+\sum_{s=0}^{\infty}\frac{(2/\pi^{2})S_{12}
\xi_{2,z}^{2}}{\left(  s+1/2+(2/\pi^{2})(\xi_{2,z}p_{z})^{2}\right)  ^{2}
}\right) \nonumber\\
\Delta_{1}^{2}(p_{z}) &  =\left(  \tau-\xi_{1,z}^{2}p_{z}^{2}-S_{12}g\left[
(2/\pi^{2})\xi_{2,z}^{2}p_{z}^{2}\right]  \right)  /b.\nonumber
\end{align}
In the linear regime $j_{z}\approx(4e\tau N_{1}/b)\left(
\xi_{1,z}^{2} +S_{12}\xi_{2,z}^{2}\right)  p_{z}$. This means that
in the whole range $(T_{c}-T)/T_{c}\ll1$ the $z$ component of the
London length
is given by the GL formula (\ref{LambdaGLApprox}). In conventional
superconductors the dependence $j_{z}(p_{z})$ is nonmonotonic and
its maximum gives the well-known GL result for depairing current,
$j_{dp}=c\Phi_0/(12\sqrt{3}\pi^2\lambda^2\xi)\propto \tau^{3/2}$
for $\tau\rightarrow0$.\cite{Tinkham} In our case the situation is
different. The amplitude of the order parameter is suppressed at
$p_{z}\sim1/\xi_{z}(T)$. However, in the region
$\xi_{z}(T)\ll\xi_{2,z}$ the dependence $j_{z}\left( p_{z}\right)
$ becomes nonlinear at much smaller $p_{z}$,
$p_{z}\sim1/\xi_{2,z}$. The shape of this dependence is determined
by the parameter $S_{r}=S_{12}\xi_{2,z} ^{2}/\xi_{1,z}^{2}$. The
dependencies $j_{z}(p_{z})$ for $S_{r}=6$ and different
temperatures are plotted in the left panel of Fig.\ 2. For large
values of $S_{r} $ the dependence $j_{z}(p_{z})$ has \emph{two
maxima} within some temperature range, where first (second)
maximum corresponds to the suppression of supercurrent in the
$\pi$-($\sigma$-)band. At low temperatures the depairing current
$j_{dp}$ is given by the second maximum and is determined mainly
by the $\sigma$-band. At certain temperature near $T_{c}$ global
maximum switches to the first maximum (see Fig.\ 2). The
temperature dependence of $j_{dp}$ has a kink at this temperature
(see the right panel of Fig.\ 2). For $S_{r}>6$ the local maximum
of $j_{z}\left( p_{z}\right) $ at $p_{z}\sim1/\xi_{2,z}$ exists
even in the limit $\xi _{z}(T)\ll\xi_{2,z}$.

As another example, we compute from Eqs.\ (\ref{Eqs_z})
the in-plane upper critical field near $T_{c}$, $H_{c2,a}(T)$.
Experiment \cite{Hc2Exp} shows strong upward curvature of
$H_{c2,a}(T)$, leading to the temperature dependent anisotropy
factor. Microscopic calculations reproduce this feature, both in
clean \cite{Hc2Clean} and dirty \cite{GK-Hc2} cases, but require
rather heavy numerical computations. Our model allows to trace
origin of the upward curvature in a simple way.
%
Selecting the gauge $A_{z}=Hx$ and introducing reduced variables
$h=H/H_{c2}^{(1)}$ with $H_{c2}^{(1)}\equiv \Phi_0/(2\pi
\xi_{1,x}\xi_{1,z})$, $x\rightarrow \sqrt{h}x/\xi_{1,x}$,
$r_{z}=\mathcal{D}_{2,z}/\mathcal{D}_{1,z}$, we write the linear
equation for determination of the upper critical field,
$h=H_{c2}/H_{c2}^{(1)}$, as
\begin{subequations}
\begin{align}
-\frac{S_{12}}{h}\sum_{s=0}^{\infty}\left(
\frac{\Delta_{1}}{s+1/2}\!-\!f_{s}\right)
\!-\!\nabla_{x}^{2}\Delta_{1}\!+\!x ^{2}\Delta_{1}
&\!=\!\frac{\tau}{h}\Delta_{1}  ,\\
(s+1/2)f_{s}+\frac{2}{\pi^{2}}r_{z}hx^{2}f_{s} & \!=\!\Delta_{1}.
\end{align}
\end{subequations}
Excluding $f_{s}$, we obtain the Schroedinger
equation for $\Delta_{1}$ with non-parabolic potential
\begin{equation}
-\nabla_{x}^{2}\Delta_{1}+\left( x^{2}+\frac{S_{12}}{h}\ g\left[
\frac{2r_{z} hx^{2}}{\pi^{2}}\right] \right)  \Delta_{1}
=\frac{\tau}{h}\Delta_{1}
\end{equation}
Only in the limit $h\ll\sqrt{1+S_{12}r_{z}}/r_{z}\ll1$ this
equation reduces to the usual oscillator equation. In this limit,
using expansion $g(u)\approx (\pi^2/2)u$, we reproduce the GL\
result, $h_{c2}=\tau/\sqrt{1+S_{12}r_{z}}$. The inequality
$h_{c2}\ll\sqrt{1+S_{12}r_{z} }/r_{z}$ reproduces criterion
(\ref{GLcriterion}) for the validity of the GL theory. In the
opposite limit, $\sqrt{1+S_{12}r_{z}}/r_{z}\ll h\ll1$, one can use
the asymptotics $g(u)\approx\ln\left( u\right)  +\gamma_E+2\ln 2$
for $u\gg 1$ with $\gamma_E\approx 0.577 $ being the Euler
constant, and obtain
\begin{align*}
&  -\nabla_{x}^{2}\Delta_{1}+\left( x^{2}+\frac{S_{12}}{h}\ln(
x^{2})  \right)  \Delta_{1}=\alpha(h)\Delta_{1}\\
&  \tau=\alpha(h)h+S_{12}\left(  \ln\left[ \frac{8hr_{z}
}{\pi^{2}}\right]  +\gamma_{E}\right)
\end{align*}
This gives the following equation for the upper critical field
\[
h_{c2}+S_{12}\ln\left[  Ch_{c2}r_{z}\right]  =\tau
\]
with $C\sim1$ ($C\approx(8/\pi^{2})\exp\left[  \left\langle
\ln\left( x^{2}\right)  \right\rangle +\gamma_{E}\right]
=2/\pi^{2}$ for $h\gg S_{12}$). In this limit the $\pi$-band gives
only small logarithmic correction to the upper critical field. As
we can see, the upper critical field has a strong upward curvature
in a narrow region neat $T_{c}$: the slope $dh_{c2}/d\tau$ changes
from $1/\sqrt{1+S_{12}r_{z}}$ to 1 near $\tau=S_{12}+1/r_{z}$, in
agreement with microscopic calculations and experiment.

In conclusion, we demonstrated that the properties of magnesium
diboride are not described by the anisotropic GL theory. We
derived a simple model which replaces this theory in the vicinity
of $T_c$, and explored some consequences of this model.

We would like to thank U.\ Welp for critical reading of the
manuscript. This work was supported by the U.S. DOE, Office of
Science, under contract \# W-31-109-ENG-38.


\begin{thebibliography}{99}
\bibitem{criterion}In the region $\xi_z(T)<\xi_{2,z}$ equations
(\ref{minEq}) are quantitatively correct until
$W_2>\ln(\xi_{2,z}^2/\xi_z(T)^2)$.
\end{thebibliography}
\end{document}